\documentstyle[pre,aps,preprint,epsf]{revtex}
\tightenlines
\begin{document}
\draft

\title{Fracture driven by a Thermal Gradient}

\author{Oscar Pla\cite{email}}
\address{
Instituto de Ciencia de Materiales, C.S.I.C., \\ Universidad Aut\'onoma
de Madrid C-III, E-28049 Madrid, Spain.}

\maketitle
\begin{abstract}
Motivated by recent experiments by Yuse and Sano ({\sl Nature}, {\bf 362},
329 (1993)),
we propose a discrete model of linear springs for studying fracture in
thin and elastically isotropic brittle films. The method enables us to
draw a map of the stresses in the material. Cracks
generated by the model, imposing a moving thermal gradient in the material,
can branch or wiggle depending on the driving parameters.
The results may be used to compare with other recent theoretical work, or
to design future experiments.
\end{abstract}


\narrowtext
\section{INTRODUCTION.}

The study of brittle fracture is of particular technological interest for
understanding the properties of ceramic materials. By {\sl ceramic} we
understand covalent ionic materials, including also glasses, policrystaline
aggregates, minerals and even composites\cite{Lawn}.
Studies in the statistical physics and engineering aspects of the problem
\cite{NATO,SA} have been carried out, and though some proposals have been
made, a comprehensive and general understanding of the mechanisms leading
to fracture is still missing.
The fundamental reason for this apparent lack of progress may be that
experimental work in the study of these processes is very difficult because
of the great quantity of parameters that seem to have impact in the resulting
shape or velocity of the crack. Recently Yuse and Sano carried out an
experiment\cite{YS} in which they could control the fracture pattern as
function of the external imposed stresses. It is sketched if Fig.\ 1. A thin
rectangular plate of glass with an initial notch is pulled at a velocity $v$
inside a region of width 2$\xi$, where there is a change in the temperature
from $T_h$ to $T_c$. As function of increasing $v$, and also
$\Delta T$=$T_h-T_c$, several
phases may be found: first, for low values of the control parameters, no
fracture is propagated. If the control parameter, i.e. $v$, is increased, a
straight crack is generated. If it is again increased, a wiggling pattern
similar to that of Fig.\ 1 appears. And, finally, for high enough values of
$v$ branching occurs, and more than one crack propagates in the system.

This experiment has inspired directly analytical\cite{MM,SSN} as well as
numerical\cite{H1,H2} work. Recent numerical simulations\cite{ABRR} also
show this oscillatory instability when the system is driven, not by a thermal
gradient, but by an increasing strain in the borders.

The purpose of this paper is to introduce a discrete model that has the right
long wave behavior of the continuum elasticity theory, and to check its
predictions of wiggling cracks in the case of a thermally driven fracture. Here
I will thus try to present the tool. Quantitative comparisons against
analytical models and experiment will be the subject of future publications.

The basic physical ideas that inspire the model developed here are that
brittle materials are elastic, in the strict sense that displacements are
proportional to applied stresses, until they fail. Also, since the velocity
of the crack propagation is imposed by the motion of the temperature change,
and it is much smaller than the sound velocity in the material, cuasistatic
calculations are good candidates to give a right description of the problem.

In section II I will describe the model, expose the fracture processes, and
see how the model reproduces qualitatively the thermally driven fracture
scenario. In section III I will show how the stress profiles in the system
can be calculated. Section IV will be dedicated to study the
dependence of the different parameters when we want to produce larger
systems (or better discretizations). Finally,
in section V, we will present some conclusions and paths for future works
and applications of the model.

\section{MODEL AND SIMULATIONS.}

\subsection{Discretization.}

The continuum elastic description of a two dimensional material that is in the
presence of a temperature field, $T$, is given by the free energy\cite{LL}
\begin{equation}\label{eq:elast}
{\cal F}_e =\frac{1}{2}(\lambda +\mu)(\varepsilon_{xx}+\varepsilon_{yy})^2
+\mu(2\varepsilon_{xy}^2+\frac{1}{2}(\varepsilon_{xx}-\varepsilon_{yy})^2)
-(\lambda +\mu)\alpha T(\varepsilon_{xx}+\varepsilon_{yy}),
\end{equation}
where $\varepsilon_{ij}=(\partial_iu_j+\partial_ju_i)/2$ are the components of
the strain tensor ($u_i$ are the local deformations of the material in the
$i$ direction), $\lambda$ and $\mu$ the Lam\'e coefficients, and $\alpha$
the thermal expansion.

As simple microscopic model we can consider a system of hookean springs
described by the free energy
\begin{equation}\label{eq:springs}
{\cal F}_s = \frac{1}{2}\sum_{ij}k_{ij}\left[({\bf u}_j-{\bf u}_i)\cdot
\hat{r}_{ij}\right]^2
-\sum_{ij}k_{ij}W_i\left({\bf u}_j-{\bf u}_i\right)\cdot\hat{r}_{ij},
\end{equation}
where ${\bf u}_i$ is the displacement of node $i$ from its equilibrium
position, $\hat{r}_{ij}$ is the unitary vector that points from node $i$
to node $j$, $k_{ij}$ is the force constant of the spring that joins
$i$th and $j$th nodes, and $W_i$ is the scalar field that includes temperature,
compression modulus, and thermal expansion at node $i$\cite{comp:H}.
If the nodes are placed in a triangular lattice, and
links are made all with the same force constant and only between nearest
neighbors, the long wavelength behavior of Eq.\ \ref{eq:springs}
is the same as that described by the continuum Eq.\ \ref{eq:elast}, when the
temperature is constant in the material, and with $\lambda$=$\mu$.

Let us introduce a reference system in terms of the axis of the triangular
lattice, making the following changes of variables: $u=(2/\sqrt{3})x$ and
$v=y+(1/\sqrt{3})x$ for the coordinates,
and $u_u=(\sqrt{3}/2)u_x-(1/2)u_y$, $u_v=u_y$, and $u_w=u_u+u_v$ for the three
vector displacements in the triangular lattice\cite{note1}. We will also
take the vectors $\hat{r}_{ij}$ as vectors in the triangular lattice, making
the equations of motion linear in the displacements.

The free energy of the model can be written as
\begin{equation}\label{eq:springs.c}
{\cal F}_s = \frac{1}{2}\sum_n\left((\partial_uu_u)^2+(\partial_vu_v)^2+
(\partial_wu_w)^2\right)
-2\sum_nW_n\left(\partial_uu_u+\partial_vu_v+\partial_wu_w\right),
\end{equation}
where $n$ now sweeps all the nodes in the system, and the derivatives are
evaluated as differences between neighbor nodes when there are links between
them ($k_{ij}$ takes values 0 (1) for missing (present) springs).

\subsection{Temperature.}

If the temperature drop is being displaced at a velocity $v$ in the
$y$ direction, the thermal diffusion equation
\begin{equation}
\frac{\partial T}{\partial t}=\kappa\nabla^2T,
\end{equation}
accepts solutions that depend only on $\psi=y-vt$ in the form:
\begin{equation}\label{eq:temp}
T(\psi)=\frac{T_h-T_c}{2}+\frac{T_h+T_c}{2}\frac{\cosh{{\rm v}\xi}
-\exp{{\rm v}\psi}}{\sinh{{\rm v}\xi}},
\end{equation}
where v$\equiv v/\kappa$ is the inverse of the temperature length,
and $2\xi$ is the distance between the hot and
the cold temperature baths. Function $T$ is also sketched in Fig.\ 1.
Notice that its derivative is not continuous at $T=T_c$, where the fastest
changes in $T$ are found, and, for that reason, larger stresses are located.

\subsection{Initial conditions and relaxation.}

To perform a simulation the nodes are placed in a rombus of dimensions
$N_u\times N_v$, in units of the node-node distance. Then the corners are cut
to make a rectangular sample of dimensions $(\sqrt{3}/2)N_u\times N_v$-$N_u/2$.
A small notch cutting bonds at the edge simulate the experimental initial
conditions.
Now the temperature drop is introduced by applying $W$, following Eq.\
\ref{eq:temp}, with $y_0\equiv vt$ far bellow the system. The set of nodes
displacements, $\{u\}$, is
then relaxed by the Conjugate Gradients Method\cite{numrec} until forces
(elastic and thermal) at each node are balanced.

\subsection{Fracture.}

Once the system is relaxed, springs whose stress (including difference in
displacements and average $W$ in the connected nodes) is more than some
threshold
value, $\sigma_{th}$, are allowed to break. Since pure deterministic breakdown
can give rise to problems in the decision of which spring to
break\cite{H1}, we will make use of probabilistic breakdown following
Louis and Guinea model\cite{LGM}, and
considering four neighbors of the already broken springs\cite{MLSLG}. In
this model each spring has a probability of breaking proportional to
a power $\eta$ of its stress. If $\eta$ is large enough (I will take in this
work $\eta$=10), the method is, for all practical purposes,
deterministic, except in those cases close to branching where
there are two or several paths to which fracture may develop.
If after the relaxation there are no springs with stresses larger than the
threshold, the temperature field is displaced in the positive $y$-direction,
$y_0$ $\rightarrow$ $y_0$+$\Delta y$, and the system is relaxed again. If
breakdown occurs, the system is also relaxed without the broken bond and
stresses recalculated without moving the temperature profile. The simulation
stops when the crack gets close to the upper limit of the system.

\subsection{Results.}

As discussed in previously developed discrete models\cite{H1}, the different
shapes of cracks do not depend on the values of $W_h$ and $W_c$,
but on its difference, $\Delta W$. In particular, the model presented here,
as compression or dilation generate the same stressed state, depends only
on the absolute value of $\Delta W$. The units of $\Delta W$ are the same
as $\sigma_{th}$, and properties are only dependent on $\Delta W/\sigma_{th}$.
We will thus set $\sigma_{th}=1$, and discuss its possible changes in
section IV.

Several simulations with $\xi=6$, and system size 30$\times$100 are shown
in Fig.\ \ref{f:phase}. The model reproduces qualitatively well the
phenomenology of thermally driven fracture described in section I.
In Fig.\ 2 can be found several interesting configurations, ranging from
branching in the middle of the plate, some time after the crack seems
to be a wiggling one, to situations in which wiggling and straight cracks
run parallel. In fact, it seems that, with enough CPU time, besides the
line transitions from straight to wiggling, and from wiggling to branching,
a line transition can be defined for any branching level, or even for
transitions from two straight cracks to one wiggling and one straight.

This model has not the invariance reported in Ref. \cite{H1} with
respect of width$\cdot$v, or, in other words, that the effective thermal
lenght, sets the units of the width of the system. This is not the case
here because we have considered another parameter, that is, the distance
between the two applied temperature baths, 2$\xi$. As a consecuence, the
temperature as function of distance is an exponential and not an hyperbolic
tangent.

\section{FIELD OF STRESSES.}

Stresses are the derivatives of the free energy with respect
to the strains, $\sigma_{ij}=\partial{\cal F}_s/\partial u_{ij}$,
\begin{mathletters}
\begin{eqnarray}
\sigma_{xx} & = & \frac{3}{4}\left( \partial_uu_u+\partial_wu_w \right) -3W, \\
\sigma_{xy} & = & \frac{\sqrt{3}}{4}\left(\partial_wu_w-\partial_uu_u\right),\\
\sigma_{yy} & = & \frac{1}{4}\left( \partial_uu_u+\partial_wu_w \right)
+ \partial_vu_v-3W.
\end{eqnarray}
\end{mathletters}

The same equations, but for a constant factor, would have been obtained if
instead of Eq.\ \ref{eq:springs.c}, Eq.\ \ref{eq:elast} had been used. This
is a result that enhances the goodities of the triangular symmetry in the
description of homogeneous media. Eqs.\ 6 are evaluated at each node of the
system, considering the partial derivatives as finite differences of the
displacements of the neighbor connected nodes.

To illustrate the capabilities of the method, two different stress conditions
that lead to a straight crack are shown in Fig.\ \ref{f:plane}. Both system
sizes are 30$\times$100. The shading correspond to dividing the set of stresses
in 10 greys, and putting the same amount of points in each color. This is
done to see more clearly the symmetries in each component of the stress tensor,
and the differences between the two drivings. The frames are taken when the
crack tip is in the middle of the plate. The stress driven plate is simulated
by applying forces at each boundary node such that $\sigma_{xx}=\sigma_0$,
and $\sigma_{xy}=\sigma_{yy}=0$, when there are no springs cut from the plate.
Temperature now is constant (zero). For the stress driven plate, the stress
of the crack tip grows as crack evolves, where probably the cuasistatic
description used here fails due to the acceleration of fracture. Nevertheless,
qualitatively, the stress profiles shown in Fig.\ \ref{f:plane} must
remain the same in a dynamic description of the problem.

As seen from Fig.\ \ref{f:plane} the symmetries in the stress tensor for the
thermally driven fracture are more complicated than for the stress driven
one. Symmetries of different stresses that can be also found by analytically
solving the equations for the Airy function\cite{vincent}, obtaining results
consistent with the simulations.

To take a closer look to the crack tip divergence, sections along the middle
of the plate are drawn in Fig.\ 4, where $\sigma_{xy}$ is not plotted, as is,
by symmetry, zero, or, numerically, very close to zero due to effects of the
small lattices considered. Unfortunately the divergence of the stresses
near the crack tip consist only of a couple of data points that are not
enough to determine the exponents of the divergence.

\section{SIZE DEPENDENCE.}

One of the objections that may be argued about last section is that the
size of the cracks shown was too small to get any detail of questions
of fundamental importance like stress divergence at the tips, or even the
location of the tip itself. As we are treating a model that is but a mere
discretization of the elastic equations, if all the lengths of a given
simulation (system size, $\xi$, and $l_T=1/{\rm v}$), where scaled by a factor,
the new simulation would reproduce, at larger scale, the patterns obtained
in the small scale one (and much less CPU consuming one). But be forgot
to rescale the stress threshold $\sigma_{th}$, that has really units of
stress divided by length.

To guess this scaling we will follow an argument
also used in effective medium theories for elastic percolation\cite{FTG}.
We can apply a force of modulus $f$, pulling two nodes of the triangular
lattice apart. It would produce a displacement proportional to $f$, and
the proportionality constant is the effective spring constant in the system,
$1/a^*$, due to all possible connections of the two nodes through the infinite
triangular lattice\cite{dinamic}. As $\sigma_{th}$ takes in account only the
direct connection between neighbor nodes, the contribution of the medium must
be subtracted. The expression for the threshold stress, when the lengths of
the system are multiplied by $n$, is given, in this approximation, by
\begin{equation}\label{eq:sigmat}
\sigma_{th}^{(n)}=\frac{n}{a_n^*}-\left(\frac{1}{a_1^*}-1\right),
\end{equation}
where $a_n^*$ is calculated putting the force $f$ $n$ nodes apart in one
of the directions of the triangular lattice.

Eq.\ \ref{eq:sigmat} turns out to be only a good approximation (really
an upper bound) to the value that rescales the system to find the same
crack patterns reproduced. Fig.\ 5 shows such rescaling for one of the
cracks shown in Fig.\ 2. The shape (wavelenth and amplitude) are the
same for the three discretizations, but the threshold stresses are not
exactly those given by Eq.\ \ref{eq:sigmat} (changes in $\sigma_{th}$ are
translated to changes in $\Delta W$ since in the simulations the convention
is that $\sigma_{th}=1$). For size 60$\times$200 (90$\times$300),
$\sigma_{th}=1.452(1.8)$ in Fig.\ 5, and from Eq.\ \ref{eq:sigmat} we would
have that $\sigma_{th}=1.475(1.952)$, values that lead to a straight crack in
the simulations.

Thus, the theoretical values for $\sigma_{th}$ obtained by this method are
not the right ones\cite{improve}, but provide a good estimation that can save
time when
we want to have larger systems and we do not want to sweep the hole parameter
space involving large simulations. As a reference, the 90$\times$300 crack
presented here took 5 days cpu time in an alpha machine.

\section{CONCLUSIONS.}

As conclusions, we have seen that hookean springs placed in a triangular
lattice, and coupled to the temperature defined at each node, provides
a discrete model for simulating brittle materials, in agreement with the
behavior of continuum elasticity theory. Simulations reproduce qualitatively
well experiments, but still, a simple description from a theoretical
point of view is missing. The simulation framework presented here can
provide a valuable tool to check new proposals for explaining fracture
phenomena. Some possible extensions of this work follow.

Here we have analyzed the case in which all the springs have the same force
constant, or in the elastic language $\lambda=\mu$. This is not the case
for most materials. In order to simulate materials with different ratios in
the Lam\'e coefficients a $\sqrt{3}$$\times$$\sqrt{3}$ reconstruction of the
triangular lattice with two different force constants may be used\cite{recons}.

It is also tempting to think of this thermally activated cracking effect as
a basis of a machinery that could microfabricate all kinds of devices. By
passing the glass film in one direction through the thermal gradient, and
then backwards in the other, and also modifying the intensity of the
temperature change, a large variety of forms can be achieved, and all with
relatively high accuracy.

The extension of the calculations done in this work to two dimensional disks
is not difficult. The question is if this kind of fracture can be accomplished
experimentally.

Another problem of technological interest for which very similar
techniques to those employed here is stress corrosion cracking. In this
case a poisoning agent plays the role of temperature by increasing locally
the volume of the material (the scalar field that couples to elastic
displacements is concentration).

\acknowledgments

I am in debt with Paco Guinea for introducing me to the subject, and for his
encouragement and discussions through this work. I also acknowledge financial
support from  the spanish CICYT Grant MAT94-0982-C02-02, and useful discussions
with Enrique Louis and Shu-zhu Zhang.

\begin{figure}
\caption{On the left, schematic representation of Yuse and Sano experiment.
The right part shows the temperature profile in the system (see text).}
\label{f:exp}
\end{figure}
\begin{figure}
\caption{Simplified phase space showing straight, oscillating and
several levels of branching for systems of size 30$\times$100 and $\xi=6$.}
\label{f:phase}
\end{figure}
\begin{figure}
\caption{Stresses in the plate, for two different stress conditions that
give rise to a straight crack: in the upper part thermally driven with
$\Delta W=1.8$, ${\rm v}=1$, and $\xi=6$; and in the lower part uniform
uniaxial
stress in the $x$-axis ($\sigma_0=1.5$). Both profiles were taken when
the crack is at the middle of the plate.
System sizes are, for both cases, 30$\times$100.}
\label{f:plane}
\end{figure}
\begin{figure}
\caption{Cross section of the stresses shown in Fig.\ 3 along the crack.}
\end{figure}
\begin{figure}
\caption{One of the wiggling crack patterns of Fig.\ 2 ($\Delta W=1.8$, and
${\rm v}=3$), and scaling it for twice and
three times the involved lengths. For twice $\Delta W=1.24$, and for three
times $\Delta W=1.0$.}
\end{figure}

\begin{references}
\bibitem{note1}
Note that this reference system is not orthogonal, but the election makes
easier to work with triangular lattices.
\bibitem{dinamic}
The explicit calculation is done by expressing the fourier transform of
the displacements as product of minus the dynamical matrix times the
fourier transform of the applied forces, $u_k=-D(k)F_k$. And back to real
space integrating over the Brillouin zone of the triangular lattice. For
more details see \protect\cite{FTG}.
\bibitem{improve}
The theoretical values of $\sigma_{th}$ can get closer to those found
euristically in the simulations by including the finite size of the plates
on the modification of the Brilluin zone, but still do give straight cracks
for large change of system size. A more detailed calculation should involve all
possible paths of stresses throug the system, and may be the implicit
consideration of the temperature.
\end{references}
\end{document}